\documentclass[12pt]{article}
\usepackage{caption,color,fullpage,graphicx,natbib,subcaption}

\begin{document}

\title{Discussion of: ``A Bayesian information criterion for singular models''}
\author{N. Friel$^{1,4}$, J. P. McKeone$^2$, C. J. Oates$^{3,5}$, A. N. Pettitt$^{2,5}$\\ \small
$^1$ School of Mathematics and Statistics, University College Dublin, Ireland \\ \small
$^2$ School of Mathematical Sciences, Queensland University of Technology, Australia \\ \small
$^3$ School of Mathematical and Physical Sciences, University of Technology Sydney, Australia \\ \small
$^4$ Insight: Centre for Data Analytics, Ireland \\ \small
$^5$ ARC Centre of Excellence for Mathematical and Statistical Frontiers, Australia }
\date{}
\maketitle

The authors should be congratulated on their thought-provoking contribution in \cite{Drton2017}.

Our discussion focuses on the widely applicable Bayesian information criterion \citep[WBIC;][]{Watanabe2013}, an approximation to the model evidence that is valid for (both non-singular and) singular models.
The WBIC combines approximation and computation; its evaluation requires Monte Carlo (MC) but approximation is used to reduce this computational cost  compared to ``exact'' MC methods \citep[e.g.][]{Friel2008}.
In contrast to the singular Bayesian information criterion \citep[sBIC;][]{Drton2017}, analytic bounds on learning coefficients are not required for WBIC.
One can therefore implement WBIC in more general settings than sBIC if one can afford the associated MC computational cost.
A second important difference is that the prior is explicitly required for WBIC whereas it is only used implicitly in sBIC.  
The performance of WBIC can be sensitive to the prior, a basic characteristic of Bayesian model choice!  
For the galaxy data, in particular, \cite{Cameron2014} demonstrated prior sensitivity for the posterior model probabilities for the number of clusters.

\begin{figure}[t!]
\begin{subfigure}[b]{0.5\textwidth}
\includegraphics[width = \textwidth,clip,trim = 1cm 6cm 1cm 6cm]{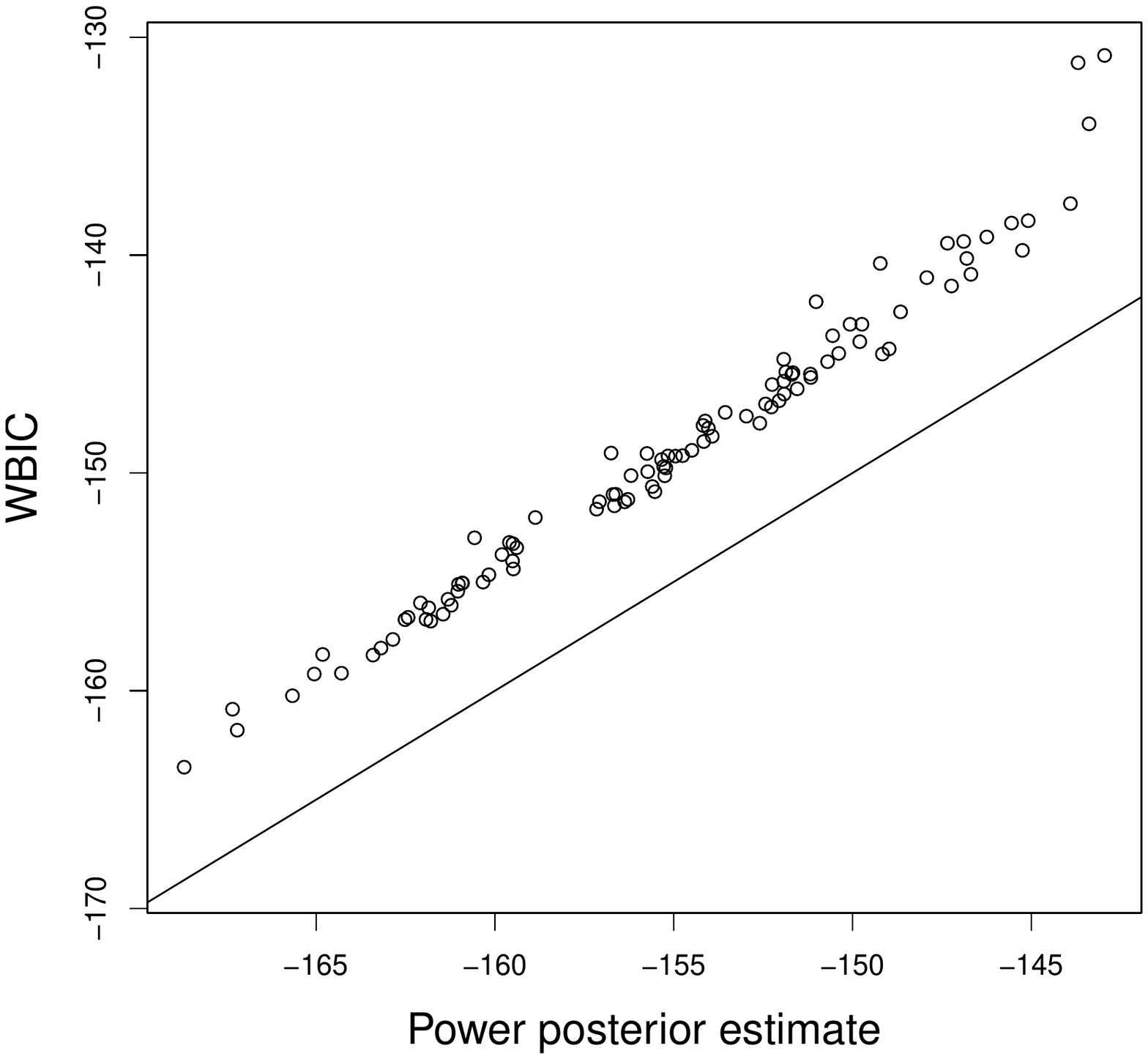}
\caption{$n=50$}
\end{subfigure}
\begin{subfigure}[b]{0.5\textwidth}
\includegraphics[width = \textwidth,clip,trim = 1cm 6cm 1cm 6cm]{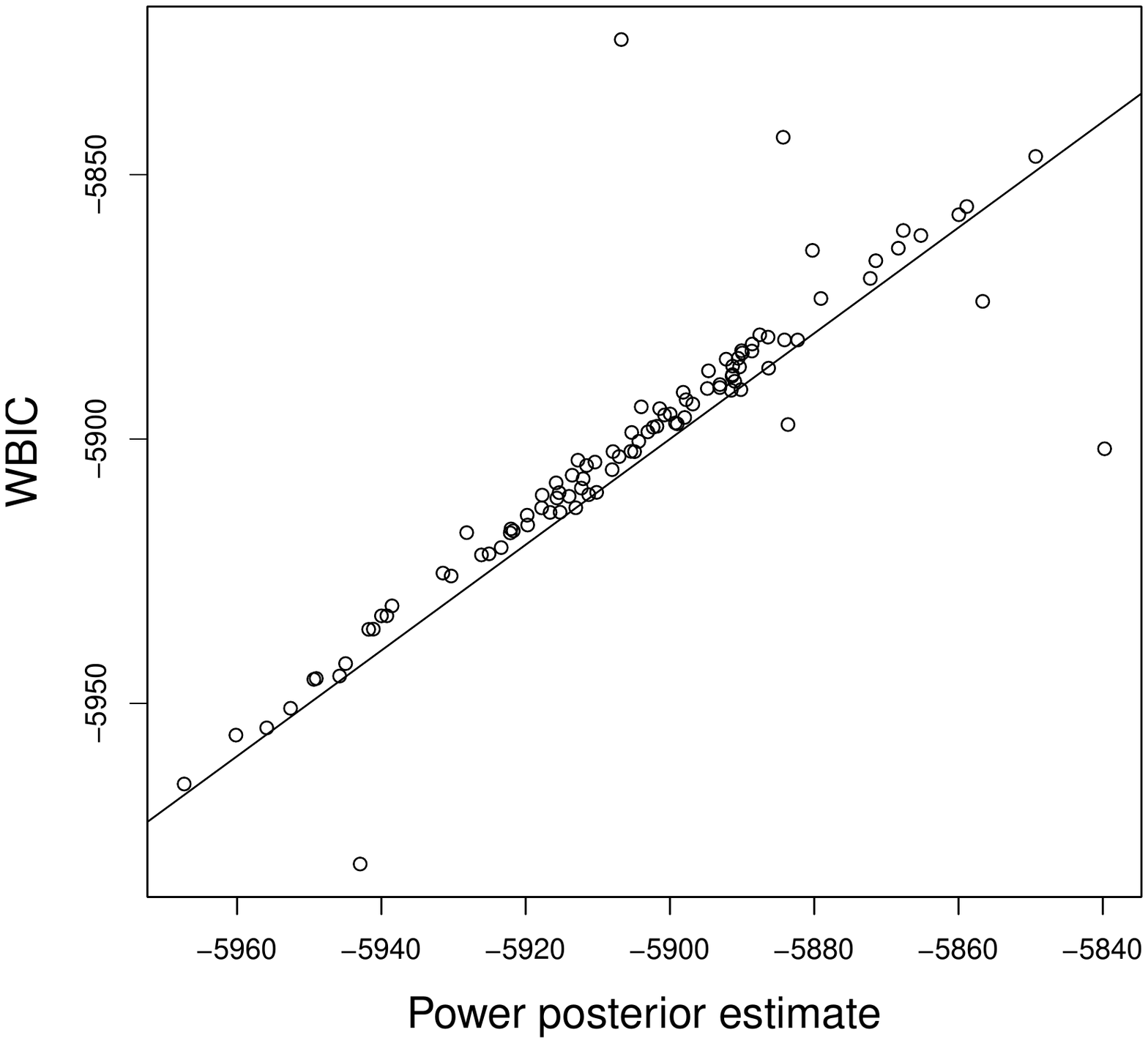}
\caption{$n=1000$}
\end{subfigure}
\caption{Finite Gaussian mixture model: WBIC against the power posterior estimate of the exact model evidence. (a) Sample of size $n = 50$. (b) Sample of size $n = 1000$. \citep[Reproduced from][each point represents an independent dataset (100 in total)]{Friel2016}.
}
\label{fig:Friel2016}
\end{figure}

It is demonstrated in Sec. 5.1 (rank selection example) that performance of WBIC is inferior to that of sBIC. 
The reason for this difference is not discussed. 
Here we complement these results with our own, which show that for Gaussian mixture models (GMMs), WBIC tends to over-estimate model evidence for GMMs.
This is, of course, particularly clear when the number $n$ of data is small since WBIC is an asympotic approximation; see Fig. \ref{fig:Friel2016}.
Results in Sec. 6.2 do not indicate whether sBIC also over-estimates model evidence for GMMs, but Fig. 4 (factor analysis) suggests that sBIC is not over-confident in a different class of problem. 
It would be interesting to see whether sBIC has an advantage over WBIC in the GMM example in terms of avoiding over-confident approximation of model evidence terms.

The sophistication of modern statistical models demands intelligent approximation methods.
Tractable approximations to model evidence are an important research goal, whether based on asymptotic results or on efficient numerical approximation methods. 
In particular, promising research directions include the use of approximate MC methods \citep{Alquier2016} and variance reduction techniques \citep{Oates2016}.

\bibliographystyle{plainnat}
\bibliography{bibliography}

\end{document}